# ZK-WAGON: Imperceptible Watermark for Image Generation Models using ZK-SNARKs

Aadarsh Anantha Ramakrishnan
*Department of Computer Science and Engineering*
*National Institute of Technology, Tiruchirappalli*
Tiruchirappalli, India
aadarsh.ram@gmail.com

Shubham Agarwal
*Department of Computer Science and Engineering*
*National Institute of Technology, Tiruchirappalli*
Tiruchirappalli, India
shub.agarwal2003@gmail.com

Selvanayagam S
*Department of Computer Science and Engineering*
*National Institute of Technology, Tiruchirappalli*
Tiruchirappalli, India
snsn010212@gmail.com

Kunwar Singh
*Department of Computer Science and Engineering*
*National Institute of Technology, Tiruchirappalli*
Tiruchirappalli, India
kunwar@nitt.edu

*Abstract*—As image generation models grow increasingly powerful and accessible, concerns around authenticity, ownership, and misuse of synthetic media have become critical. The ability to generate lifelike images indistinguishable from real ones introduces risks such as misinformation, deepfakes, and intellectual property violations. Traditional watermarking methods either degrade image quality, are easily removed, or require access to confidential model internals – making them unsuitable for secure and scalable deployment. We are the first to introduce ZK-WAGON, a novel system for watermarking image generation models using the Zero-Knowledge Succinct Non-Interactive Argument of Knowledge (ZK-SNARKs). Our approach enables verifiable proof of origin without exposing model weights, generation prompts, or any sensitive internal information. We propose Selective Layer ZK-Circuit Creation (SL-ZKCC), a method to selectively convert key layers of an image generation model into a circuit, reducing proof generation time significantly. Generated ZK-SNARK proofs are imperceptibly embedded into a generated image via Least Significant Bit (LSB) steganography. We demonstrate this system on both GAN and Diffusion models, providing a secure, model-agnostic pipeline for trustworthy AI image generation.

## I. INTRODUCTION

The rise of high-fidelity AI image generation models like Generative Adversarial Networks (GANs) [1] and Diffusion Models [2] has created urgent challenges regarding authenticity, ownership, and malicious use, such as deepfakes and disinformation. Conventional watermarking methods are often ineffective, as they can be visually intrusive, easily removed, or require access to proprietary model information for verification.

To address these issues, we introduce **ZK-WAGON** (**ZK**-SNARK **W**atermark for im**AG**e generati**ON** models), a novel watermarking system for image generation models using Zero-Knowledge Succinct Non-Interactive Arguments of Knowledge (ZK-SNARKs). The method enables a provable origin claim for generated images, to prove that they were generated using a specified model without revealing any sensitive internal information. The core of our system is a process called **Selective Layer ZK-Circuit Creation (SL-ZKCC)**. This converts only the most important layers of an image generation model into a ZK-SNARK circuit, which minimizes the computational cost and time required to generate proofs. For intricate models where proof generation remains resource-intensive, we acknowledge the utility of cloud-based platforms like Lilith [12]. The generated ZK-SNARK proof, which cryptographically guarantees the image's origin, is then imperceptibly embedded within the image using Least Significant Bit (LSB) steganography. During verification, users can extract this proof to confirm which model generated the image. To summarize, our key contributions are as follows:

- We introduce ZK-WAGON, a novel model-agnostic framework for watermarking image generation models using ZK-SNARKS.

- We propose SL-ZKCC, a process for converting important layers of an image generation model into a ZK-SNARK circuit.

- We use a combination of LSB steganography, hashing and compression to hide zero-knowledge proofs imperceptibly within an image, and allow third-party users to verify the origin of a generated image.

## II. RELATED WORK

### A. Watermarking Approaches for Generative Models

The rise of generative AI has led to various watermarking techniques aimed at establishing ownership, authenticity, and traceability without degrading image quality. Qiao et al. [3] embedded watermarks directly into GAN generator networks, enabling ownership verification while preserving visual fidelity, though the method was GAN-specific and relied on a verification image and key. Yang et al. [4] proposed Gaussian Shading for Diffusion Models, offering provable, lossless watermarking via latent representations, but its applicability was limited to diffusion-centric models. Lin et al. [5] developed a CycleGAN-based method using a custom loss and frozen decoder, resistant to transformations but incompatible with diffusion models and vulnerable to white-box attacks.

These approaches show significant advancements in watermarking for generative models, but they fall short in delivering mechanisms that are model agnostic, making their adaptability difficult to other image generation models.

## B. ZK-SNARKs for Model Verification

A ZK-SNARK is a type of cryptographic proof that allows computations to be proved without revealing the underlying data or logic. They are therefore well suited for applications in model inference verification where privacy and trust are paramount. Sheybani et al. [6] introduced ZKROWNN using the Groth-16 protocol for DNN ownership verification, though embedding watermarks in early layers altered model weights. Sato and Tanaka [7] used ZK-STARKs to verify DNN outputs via Merkle tree roots, but their method risked model weight exposure. Germani et al. [8] applied ZK-SNARKs and blockchain for inference verification in workplace injury claims, deploying smart contracts for validation.

These studies provide methods for privacy-preserving inference verification using ZK-SNARKS, but they do not provide methods for creating ZK proofs for image generation models. ZK-WAGON attempts to bridge this literature gap through the introduction of SL-ZKCC.

## III. METHODOLOGY

We structure our methodology into **three principal phases**:

1) The *image generation phase*, which relates to the synthesis of images from generative models.
2) The *cryptography phase*, designed for the construction of ZK-SNARK circuits and the creation of imperceptible watermarks using ZK-SNARK proofs.
3) The *verification phase*, which allows users to verify watermarked images to understand their origin.

## A. Image Generation Phase

This is the initial phase of our framework, where digital images are synthesized to serve as the main artifacts for the watermarking process that follows. To show our framework's model agnostic capability, we employ two predominant image generation model architectures: **Generative Adversarial Networks** (GANs) and **Diffusion Models**. We utilize TinyGAN [9], a computationally efficient model derived from BigGAN [19] using knowledge distillation. For diffusion-based image generation, Stable Diffusion 2.1 Base [10] is used with a slight tweak - its native Variational Autoencoder (VAE) is substituted with Tiny AutoEncoder for Stable Diffusion (TAESD) [14]. This is a lightweight, distilled version of Stable Diffusion's VAE, designed for swift decoding of latent representations into full-sized images, thus being suitable for applications that require speed. TAESD maintains a high degree of visual fidelity at the expense of slightly lower fine-grained details while being compatible with Stable Diffusion 2.1 Base. This is an improvement that dramatically reduces the inference speed without affecting high visual quality.

## B. Cryptography Phase

### 1) Zero-Knowledge Succinct Non-Interactive Argument of Knowledge (ZK-SNARKs):

ZK-SNARKs are a family of zero-knowledge proof (ZKP) systems that offer powerful guarantees. At the core of their reliability are two essential properties: **Completeness and Soundness**. The Completeness property states that if an image genuinely originates from the claimed model and the prover (e.g., the model owner) follows the protocol correctly, they can create a valid proof that an honest verifier (e.g., an image recipient) will accept; in short, legitimate claims can always be proven. By contrast, the Soundness property ensures that it is computationally hard for a malicious prover to create a proof that an honest verifier will accept.

The **Zero-Knowledge property** is very important, as it allows us to prove that an image was actually generated by a certain AI model using its legitimate (but private) weights and potentially sensitive input prompts, all without revealing private inputs. This directly addresses the core challenge of verifying origin while maintaining model confidentiality. In addition, the **Succinctness** property ensures that the cryptographic proofs are small in size, with verification being both computationally light and fast. Finally, the property of **Non-Interactiveness** is that, once a proof has been generated, it can be verified by any party with knowledge of the public parameters, without additional communication or interaction with the original prover. Collectively, these properties make ZK-SNARKs an exceptionally robust and well-suited foundation for building a trustworthy and privacy-preserving verification mechanism for AI-generated media.

### 2) Selective Layer ZK-Circuit Creation (SL-ZKCC):

The pretrained image generation model must now be transformed into a suitable format for cryptographic processing. A salient contribution in this step of our framework, is the **incorporation of SL-ZKCC**. This approach aims to precisely identify and select a minimal but *cryptographically sufficient subset* of the neural network's layers. It also efficiently resolves the high computational complexity associated with the generation of ZK-SNARK circuits for deep neural networks that conventionally comprise millions of parameters. For GAN models, these are usually the starting layers of the generator network that *project the latent vector* to intermediate feature representations. In the context of diffusion models, a latent representation in the *autoencoder's bottleneck layer* is selected. This intentional pruning of the model drastically minimizes the size and complexity of the generated ZK circuit, without losing cryptographic integrity. The selected layers are then saved in the Open Neural Network Exchange (ONNX) [13] format, whose computational graph is then translated into a Rank-1 Constraint System (R1CS) compatible with ZK-SNARK systems.

**Circuit Calibration** is subsequently executed to ensure the ZK circuit operates with correctness under the imposed cryptographic constraints. This involves generating and refining a configuration file that sets key parameters like input visibility, numerical scale, bit precision, and error tolerance. For our models, we calibrate using *10 batches* of input data.

Next, a one-time setup process generates a Structured Reference String (SRS). Using this SRS and the compiled arithmetic circuit, *two important cryptographic keys* are generated: a confidential **Proving Key** ($P_k$), maintained by the entity responsible for generating proofs, and a public **Verification Key** ($V_k$), for third-party verifiers. The entire process of circuit compilation, calibration, and key generation is managed by the EZKL toolkit.

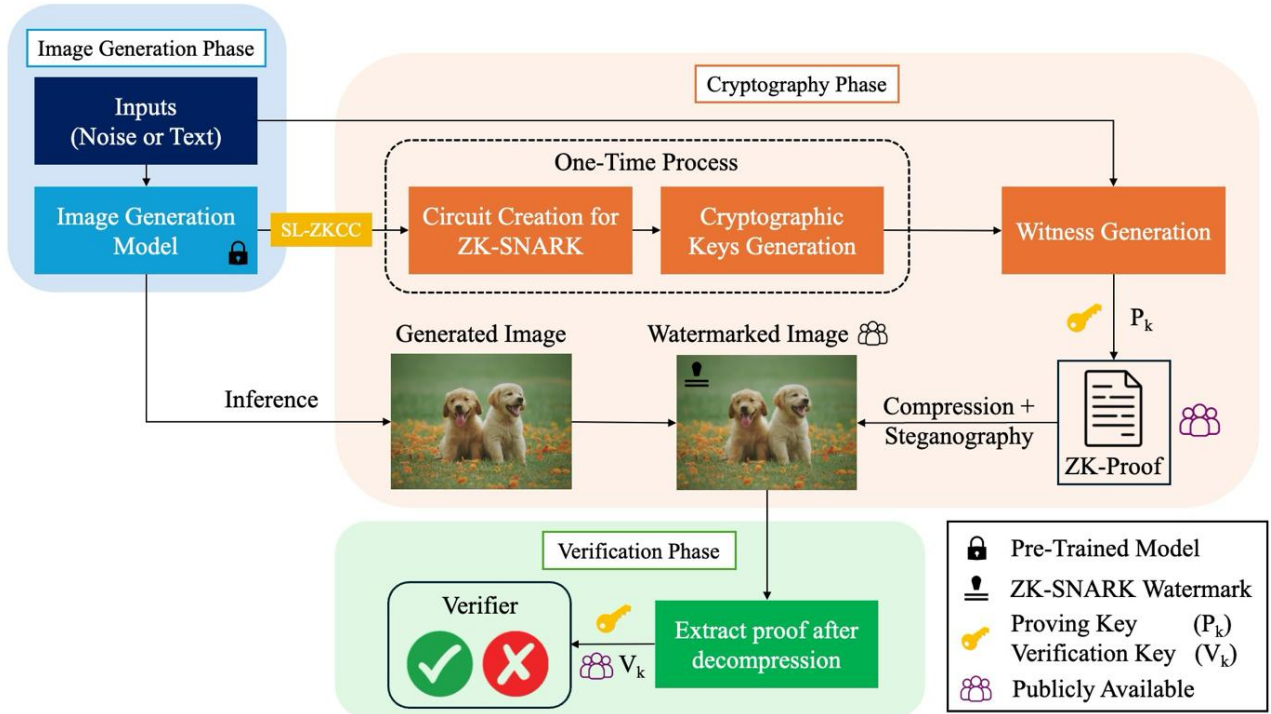

Fig. 1. Architecture Diagram of ZK-WAGON

3) **ZK-SNARK Proof Generation:**

After designing the ZK-SNARK circuit structure and the cryptographic keys needed, we generate a witness file. A *witness is a complete assignment of values* in the arithmetic circuit; this includes all private inputs (such as the initial noise vector or specific prompt embeddings), public inputs (e.g., certain model outputs or identifiers), and all intermediate computation values resulting from the selectively chosen layers (from SL-ZKCC), which together satisfy all constraints outlined in the R1CS.

In the witness preparation step, the **construction of Zero-Knowledge Proof** is initiated. In this work, we leverage the Halo2 proving system [15], an efficient ZK-SNARK scheme notable for its utilization of a transparent setup, eliminating the necessity for a trusted setup process for its SRS. The generated witness, together with the proving key ($P_k$), is passed to the Halo2 proving algorithm. The result is a compact JSON, containing the proof itself, associated public inputs, and metadata such as the proving scheme and circuit version. This proof cryptographically guarantees that the generated image originates from the given model and inputs while hiding proprietary weights and private inputs. To handle the heavy computation, we outsource proof generation to *Lilith*, a cloud platform for ZK-SNARKs.

4) **Imperceptible Watermark Creation:**

The ZK-SNARK proof will then be added to the generated image using the **Imperceptible Watermarking** pipeline. It contains several steps to securely link the proof to the image:

  a) *Perceptual Hashing and Signature Creation:*
  To link the ZK-SNARK proof and the visual content of the generated image, a 512 bit perceptual hash (using average hashing) is computed. The string form of this hash is concatenated with a fixed secret key (known only to the model owner) and then hashed with SHA-256. The resulting digital signature is embedded in the JSON file containing the ZK-SNARK proof as an *additional key–value pair*.

  b) *Proof Compression:*
  The updated JSON proof, now including the embedded signature, can still have a significant file size (~1 MB). It is losslessly compressed using GZip (uses a blend of the LZ77 algorithm [16] and Huffman coding), reducing proof size to around 100 KB, making it suitable for steganographic embedding.

  c) *Steganographic Embedding:*
  The compressed proof is invisibly embedded in the generated image via Least Significant Bit (LSB) steganography. The embedding process starts with the embedding of a 64-bit encoding of the length of the payload (in bytes). The algorithm processes the image pixels and their constituent color channels one by one, systematically altering the LSBs of these channels to encode the compressed proof bits. For a typical ~100 KB payload and a 512x512 pixel image, this process generally utilizes the 0-th LSB plane entirely and a minor portion of the 1st LSB plane. To the human eye, the watermarked image produced by this process is essentially indistinguishable from the original.

After the completion of all these steps, we have successfully created a watermarked image containing a ZK-SNARK proof. This image can now be given to the end user, who can verify the proof whenever they require.

**5) Proof Verification:**

To validate the ZK-SNARK proof ($\pi$), the watermark is initially recovered from the LSB's of the image, and then further decompressed with GZip. We then recompute the hash of the image and match it against the signature contained within the proof. This ensures that the provided image has not been tampered/modified. The verifier then checks the ZK-SNARK proof using the extracted components, as well as any necessary public inputs (e.g., hash of the model identifier, the public output hash if it was used), and the public Verification Key ($V_k$) corresponding to the ZK circuit. The verifier outputs a binary answer: *accept*, which means that the proof is correct mathematically and the computation it represents was performed correctly; or *reject*, indicating that the proof is invalid. Each verifier is tailored to a model's architecture, thus demonstrating that the image submitted by the user has indeed been produced by a specific image generation model.

## IV. IMPLEMENTATION DETAILS

*1) Image Generation and ZK Proof Workflow*

We deploy a hybrid image generation and verification system combining GAN and Diffusion models with a zero-knowledge proof (ZKP) pipeline. The GAN pipeline produces class-conditioned images with no external input; each generation yields a random but class-consistent output. The Diffusion pipeline, in contrast, takes a user-defined text prompt to synthesize images. Images are downloadable post-generation.

We **pre-deploy** the quantized ONNX model to Lilith along with a configuration file that defines the proof circuit and model parameters. When a new image is generated during runtime, we create a JSON file containing the model input for the layers selected from SL-ZKCC. This file is submitted to Lilith, which handles the full ZKP pipeline internally – computing the witness, generating the proof, and producing artifacts such as the proving key, verification key, and public inputs.

To cryptographically bind the proof to the generated image, a perceptual hash (512-bit average hash) is computed from the image and concatenated with a secret key. The SHA-256 of this string forms a digital signature, which is embedded as an additional field in the ZK proof JSON. Only the generated proof is retrieved via an API request to minimize data transfer and external dependency. Other artifacts remain accessible within the Lilith environment for command-line verification.

*2) User Application*

We have also created a full-stack application using FastAPI [17] and Next.js [18]. It supports two primary panels:

- **Generation Panel**: This panel allows users to select between a GAN or Diffusion model. Generated images are shown before and after watermarking, highlighting the imperceptibility of the embedded proof.

- **Verification Panel:** This panel enables users to upload watermarked images and verify embedded ZKPs within the image. In Fig. 3., "Proof Valid" indicates that the watermark was successfully verified and that the image is generated by a GAN.

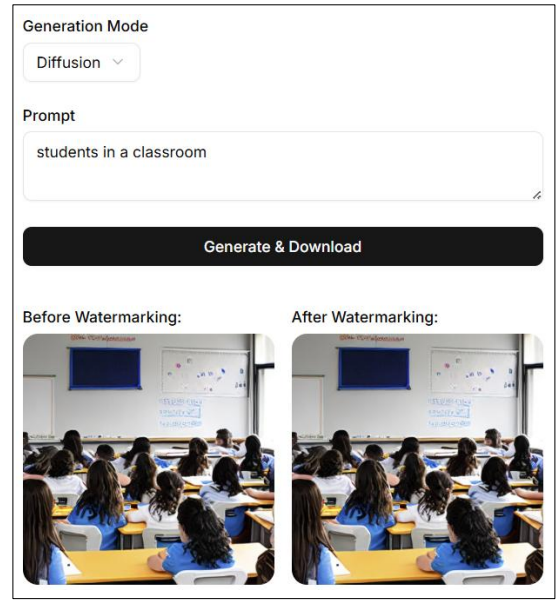

Fig. 2. Generation Panel

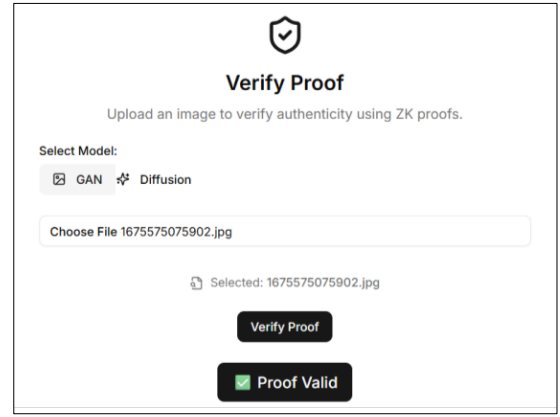

Fig. 3. Verification Panel

## V. CONCLUSION

We present a system for cryptographically verifiable image generation that incorporates ZK-SNARK proofs with both GAN and diffusion-based models. Images are generated locally, and verifiable proofs are produced using Lilith's cloud infrastructure and embedded using LSB steganography. This approach guarantees post-generation authenticity without any perceptible visual distortion. Our pipeline is model-agnostic, hardware-efficient, and end-to-end verifiable, permitting users to verify whether an image was generated by a specific model. The framework provides a scalable and tamper-evident foundation for trusted AI media generation, in the presence of increasing concerns regarding synthetic media misuse and misinformation.

## VI. ACKNOWLEDGEMENTS

We thank the EZKL team for providing us with Lilith, their enterprise compute cluster for building ZK-WAGON. Special thanks to Aashish Anantha Ramakrishnan for his contributions through insightful research discussions and proofreading.